# A perceptual hash function to store and retrieve large scale DNA sequences

Jocelyn DE GOËR DE HERVE[1-2,3], Myoung-Ah KANG[1,2], Xavier BAILLY[3],
Engelbert MEPHU NGUIFO[1,2]

[1] Clermont Université, Université Blaise Pascal, LIMOS, BP 10448, F-63000 CLERMONT-FERRAND
[2] CNRS, UMR 6158, LIMOS, F-63173 AUBIERE
[3] INRA, UR346 Épidémiologie Animale, F-63122 ST GENÈS CHAMPANELLE
Phone: 04.73.62.48.33
E-mail : jgoer@clermont.inra.fr

**ABSTRACT**

This paper proposes a novel approach for storing and retrieving massive DNA sequences. The method is based on a perceptual hash function, commonly used to determine the similarity between digital images that we adapted for DNA sequences. Perceptual hash function presented here is based on a Discrete Cosine Transform Sign Only (DCT-SO). Each nucleotide is encoded as a fixed gray level intensity pixel and the hash is calculated from its significant frequency characteristics. This results to a drastic data reduction between the sequence and the perceptual hash. Unlike cryptographic hash functions, perceptual hashes are not affected by "avalanche effect" and thus can be compared. The similarity distance between two hashes is estimated with the Hamming Distance, which is used to retrieve DNA sequences. Experiments that we conducted show that our approach is relevant for storing massive DNA sequences, and retrieve them.

## 1. INTRODUCTION

Since about ten years, the appearance of the Next Generation Sequencers (NGS) [1] allowed the production of more genomic data, genomic data, at a lower cost. The amount of data to store and analyze has experienced an exponential growth. In the data analysis pipeline, the search for similarity between DNA sequences is a basic problem common to all genomic studies. It is a necessary work during the assembly phase, to obtain a consensus sequence from several smaller ones, during the annotation tasks (to determine the biological function of a chromosome), to detect mutation, for identification of sequences or to determine a sample of biological diversity during metagenomics studies.

Regarding methods to search similarity between sequences, many algorithms and tools have been developed over the last thirty years. Most of them are derived from methods initially used for string comparison [2]. As references in the fields, we can cite, global alignment algorithms such Needleman-Wusch [3] (1970), trying to pair two nucleotide sequences at their entire length using dynamic programming techniques. Thereafter, have been developed local alignment algorithms such as BLAST [4] (1990) based on heuristics that determine common areas of several sequences. The literature also describes many indexing algorithms using hash tables (SSAHA) [5] which principle is to identify positions of different k-mers of a sequence. [6]. Compressions [7] methods for alignment and storage are also used and structures based on suffix trees [8] also allow indexing of k-mers.

Given the production of increasingly massive data from the sequencing, some of these algorithms could hardly handle the scale-up. This can be explained by their computational complexity. The complexity can be expressed as $O(n^2)$ for some of them (Needleman-Wusch), or algorithms could not be optimally parallelized. However, recent works has demonstrated some new strategies, like massive SIMD and multi-core parallelization [9]. Those methods can handle the scale-up of data production, but the principal obstacle could be the need of RAM for those algorithms that can increase exponentially, when they require loading into RAM the entire or partial raw data set to compute the tasks of comparison.

A recent publication [10] highlighted a novel technique, which uses image processing method to perform sequence alignment from Discrete Fourier Transform Phase Correlation method. This paper demonstrated the applicability of image comparison algorithms on DNA sequences, which have been previously converted into matrix of pixels. In term of sequence alignment sensibility, the proposed method has been validated, nevertheless the algorithm and its implementation was not efficient in term of execution time (hundred times slower than BLAST).

Other publications [11][12] about image processing, explore methods for indexation and comparison of images via perceptual hash functions. Overall, perceptual hashing is a set of methods and functions to generate, from the main characteristics of a document (image, sound or video), a fingerprint called hash. A hash allows a rapid identification of a document from a small amount of data. Between a document and a hash, there is a drastic reduction of the data size. However, from a hash, it is not possible to reconstruct the original document. One of the important features of perceptual hash is that the keys generated are comparable. Unlike cryptographic hash functions such MD5 [13], perceptual hashes are not affected by "avalanche effect". Thus, it is possible to determine an index of similarity between two hashes.

We performed an adaptation of a perceptual hash function based on the representation of DNA sequence as a matrix of pixels, and the application of DCT-SO [14][15] function. This technique may allow a rapid identification of query sequences against a reference sequences database. The proposed method is used to return an index of similarity between two sequences partially close or exactly the identical.

This paper is organized as follows: in the first section, we describe different steps of the hash function for DNA sequence and the comparison method. Subsequently, we describe the experimental study carried out to validate the approach and present the results.

## 2. MATERIAL AND METHODS

The method described here is based on perceptual hashing technique for digital images. We have adapted a hash function in order to use with DNA sequences and its characteristics. We therefore consider a DNA sequence as a discrete signal. It can be composed of 4 successive pseudo-random distributed states, which represent nucleotides (A, T, C, G). The hash function is based on a DCT-SO. The similarity index, which is used to compare two hashes, is calculated with the Hamming Distance [16].

### 2.1. Discrete Cosine Transform function

Discrete Cosine Transform (DCT) allows conversion of data from the spatial domain into the frequency domain. Its respective inverse functions, the Inverse Discrete Cosine Transform (iDCT) converts things back to the other way. DCT function is used to convert data into the summation of a series of cosine waves oscillating at different frequencies. It is a mathematical transformation close to the Discrete Fourier Transform (DFT) [17] but DCT involves the use of just Cosine functions and produces real coefficients, whereas DFT make use of both Sines and Cosines and require the use of complex numbers. The particularity of the DCT is that it allows expressing an image in small numbers of significant coefficients. When DCT function is applied to an image (Figure 1), a frequency representation is obtained in a matrix of coefficients. The matrix of coefficients has exactly the same dimensions as the original image. High frequencies are grouped in the upper left of the coefficients matrix. They represent the edge of an image and low frequencies represent homogeneous areas. It is therefore possible to keep only the most representative coefficients. Due to those characteristics, DCT is commonly used in image and video treatment, and it is the most important part of JPEG and MPEG formats.

$$DCT(i,j) = \frac{1}{\sqrt{2}} C(i)C(j) \sum_{x=0}^{N-1} \sum_{y=0}^{N-1} pixel(x,y) \cos\left(\frac{(2x+1)i\pi}{2N}\right) \cos\left(\frac{(2y+1)j\pi}{2N}\right)$$

**Definition of DCT for NxN matrix**: where *N* is the number of cols and lines of original image, *x* and *y* are the index of pixels in the original image, *pixel(x,y)* is the intensity value of a pixel, *i* and *j* are the indexes of the coefficients matrix and *C(i) C(j)* are the DCT coefficients

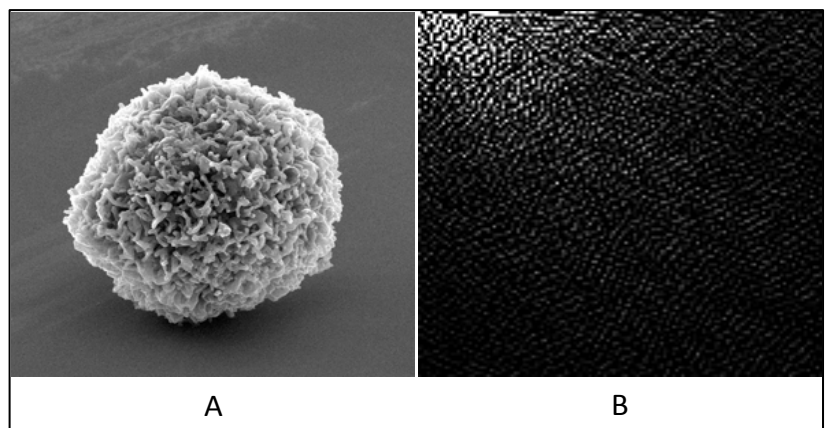

**Figure 1: A.** Original picture (295x269 pixels); **B.** Frequency representation of the image after applying a DCT function. We can observe in the top left, the grouping of high frequency, which is a characteristic of DCT.

### 2.2. DCT Sign-Only

DCT Sign-Only (DCT-SO) is an extension of the DCT. This technique is used for image comparison [18]. Its purpose is to only keep the most significant information of the image structure, in order to reduced representation cost and thus the computational time treatments. DCT-SO has similar characteristics in terms of structural image information grouping with DFT Phase. But DCT-SO is faster because it uses real coefficients unlike DFT phase, which uses complex coefficients.
DCT-SO consists in applying a *sgn()* function to DCT coefficients matrix of an image, in order to retain only signs of coefficients. This has the effect of creating binary coefficients. This is the basis of our perceptual hash algorithm. A hash corresponding to a sequence is generated from several coefficients. We keep and use the most significant coefficients to generate binary hashes.

$$\text{sgn}(DCT(i,j)) = \begin{cases} 0 & if \quad DCT(i,j) \leq 0 \\ 1 & if \quad DCT(i,j) > 0 \end{cases}$$

Definition of sign function

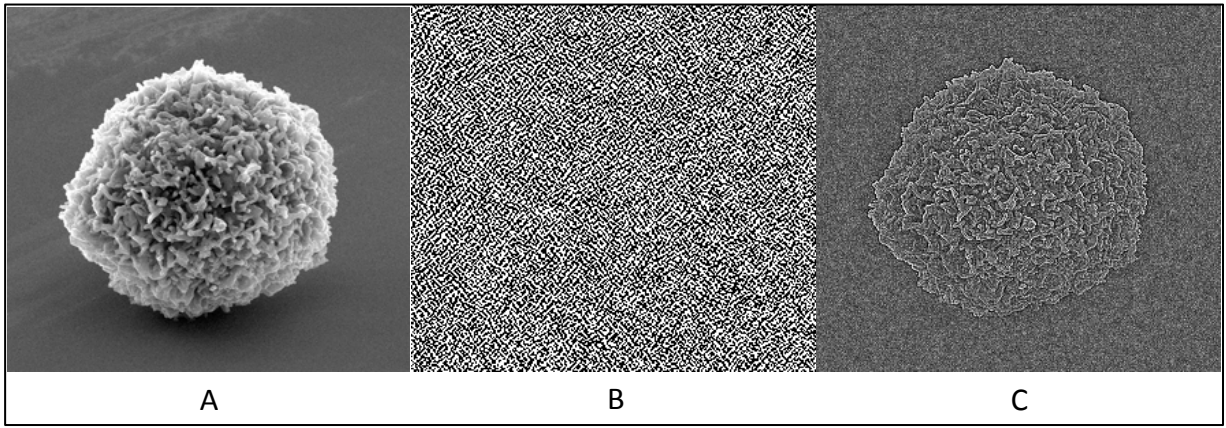

| | | |
|---|---|---|
| A | B | C |

**Figure 2**: **A.** Original image (295x269 pixels); **B.** Binaries coefficients matrix after a DCT-SO (representation in the frequency-domain); **C.** Rebuilt image from a DCT-SO coefficient matrix (representation in the special domain). The image structure is clearly visible.

### 2.3. Representation of sequences as a matrix of pixels

In order to apply the hash function, it is necessary to convert sequences into a gray scale matrix of pixels. Thus, each nucleotide is encoded by a light intensity value depending on its type. The Adenine has the intensity 63, the Thymine 127, the Cytosine 191 and the Guanine 255 (see Table 1).. These values, ranging from dark to light, were chosen because they have a constant distance.

| Nucleotid : | Value : | Pixel : |
|---|---|---|
| A | 63 | ■ |
| T | 127 | ▨ |
| C | 191 | ▨ |
| G | 256 | □ |

**Table 1**: Conversion table between nucleotides and gray scale pixel.

### 2.4. Application of perceptual hash function

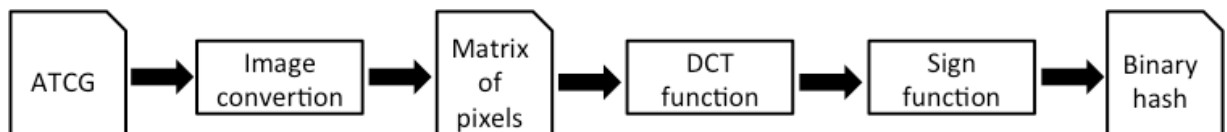

**Figure 2:** Different steps of the hash function algorithm

To compute a hash from a DNA sequence, we first convert it to a gray scale matrix of pixels (step 1) and we apply a TCD-SO function (step 2). Hashes are generated from the most relevant binary coefficients of the TCD-SO (step 3). Figure 3 illustrates our approach to hash DNA sequences.

The following example shows the different phases for hashing a DNA sequence having a size of 256 nucleotides and generating a 64-bit hash. We can notice that in this example, the hash key is 32x as smaller as the sequence in FASTA format.

```
TTAGAGTGATTTTGTACTAGCCGGACAGATACGTGAATTCTGACACCGCAAGGGCAGTTCTATC
TTAGCAAACCTTGGTTTATCTGTGGCGTTTTGTGCTACTGATGAGGATTACCTCATTCTGCACC
GAAATGACCCAGAACTAGTCATGCTGCGTCGGAAATGAACCTTCTTGAGAATTGACGATCTAAT
TTCTAGTAATCTTTCCTACTATTACCTTGCTATACCAAGCTTTCCAAACCTTCGACAGATTTGC
```

**Figure 3**: DNA sequence with a size of 256pb

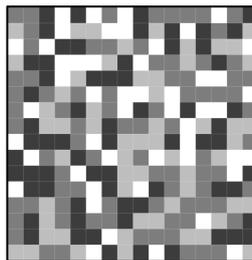

**Figure 4:** DNA sequence converted into gray scale 16x16 matrix of pixels

**Figure 5**: DCT coefficients corresponding to the matrix of pixels

```
0 0 1 1 1 1 1 0 1 0 0 1 1 0 0 0
1 0 0 1 1 0 0 0 1 0 1 1 1 0 0 0
1 0 1 1 1 0 0 0 0 0 1 0 1 1 0 0
0 0 1 0 1 1 0 0 1 0 1 0 1 0 1 0
1 0 1 0 1 0 1 0 1 1 1 0 0 0 1 1
1 1 1 0 0 0 1 1 0 0 1 1 1 0 0 0
0 0 1 1 1 0 0 0 0 1 0 1 1 1 0 1
0 1 0 1 1 1 0 1 1 0 1 1 0 1 1 0
1 0 1 1 0 1 1 0 1 1 1 1 1 1 0 1
1 1 1 1 1 1 1 0 1 0 0 0 0 0 0 0
0 0 0 0 0 0 0 0 1 1 1 0 1 1 1 0
1 1 1 0 1 1 1 0 0 1 0 1 0 1 1 0
0 1 0 1 0 1 1 0 0 0 1 0 1 0 1 0
0 1 0 0 1 0 1 0 0 1 1 1 0 0 0 0
0 1 1 1 0 0 0 0 1 1 1 1 1 0 1 1
1 1 1 1 0 1 1 1 1 1 1 1 1 1 1 1
```

**Figure 6:** After a sign() function we obtain the binary coefficients from the DCT coefficients matrix. In this example, we keep the 8x8 coefficients (in red) to form a 64 bits hash key

Hash : 00111110 10011000 10111000 00101100 10101010 11100011 00111000 01011101

**Figure 7:** 64-bits binary hash of sequence

### 2.5. Similarity estimation between two hashes

Using the Hamming Distance performs comparison between two hashes (see Figure 7).. The Hamming Distance is a mathematical distance, which expresses a sum of the differences between two sequences having the same length. The sequences can be composed of binary numbers, but also elements from other numeric or alphanumeric systems. It has a low computational complexity and it is commonly used with perceptual hash functions described in the literature. It would be possible to use other distances functions as the Leveinstein Distance [19], but this distance requires higher computing time because it allows comparison of sequences of different sizes. Hamming Distance returns an index, the more the index lower is, the more sequences are similar.

Hash 1 : 00111110 10011000 10111000 00101100 10101010 11100011 00111000 01011101
Hash 2 : 00110110 10011000 10111100 00101100 10101010 11101111 00111000 01011101

**Figure 8:** Example of Hamming Distance between two 64-bits sequences.
Hamming Distance = 4.

### 3. EXPERIMENTAL EVALUATION

#### 3.1. Theoretical evaluation by simulation

To evaluate our hash function, we performed theoretical tests by simulation. The evaluation phase consisted to perform a large number of comparisons between sequences having strictly the same length. These simulations were designed to study the statistical variability of the Hamming Distance between two sequences and thus determine the sensitivity of the similarity index based on the following parameters: the sequence length, the divergence rate and the size of hashes.

| Group | Seq. length | Hash length | Factor reduction | Divergence rate |
|---|---|---|---|---|
| A | 100 pb | 4 octets (32 bits) | 25x | 5%, 10%, 30%, 50%, 100% |
| B | 100 pb | 8 octets (64 bits) | 12,5x | 5%, 10%, 30%, 50%, 100% |
| C | 1000 pb | 4 octets (32 bits) | 250x | 5%, 10%, 30%, 50%, 100% |
| D | 1000 pb | 8 octets (64 bits) | 125x | 5%, 10%, 30%, 50%, 100% |
| E | 10000 pb | 4 octets (32 bits) | 2500x | 5%, 10%, 30%, 50%, 100% |
| F | 10000 pb | 8 octets (64 bits) | 1250x | 5%, 10%, 30%, 50%, 100% |

**Table 2**: Groups of simulations

For each group of simulations, a primary data set composed of 10 million of primary sequences was randomly generated. For each of these sequences, six different sequences were generated with strict differences rate of 5%, 10%, 20%, 30%, 50% and 100%. A hash key corresponding to each sequence was calculated. Thus, hashes corresponding to primary sequences were compared with their divergent sequences. A total of 6 million sequences have been generated to produce 7 million comparisons.

#### 3.2. Software implementation

All simulations were performed on a computer with an Intel Xeon E5-4620 @ 2.20GHz (16-Cores CPU), 128 Gb of RAM and running on Linux Ubuntu 14.04. The software implementation has been written in C++. DCT function comes from OpenCV v2.49 and random functions from the BOOST library v1.55. Whole simulations ran on 32 threads in parallel, using OpenMP directives.

### 3.3. Execution time

| Group: | Time: | Hashes per second: |
|---|---:|---:|
| A | 177s | 564 515 |
| B | 225s | 443 037 |
| C | 3 686s | 27 125 |
| D | 3 714s | 26 922 |
| E | 396 825s | 252 |
| F | 408 163s | 245 |

**Table 3**: Execution time per groups

Table 3 shows the execution time for different groups of simulations. It should be noted that a significant part (about 40%) of the execution time is due to the random generation of divergent sequences. In the same way, simulations working 64-bits hashes (Groups B, D and F) seem to be slower compared with groups evaluating 32-bits hashes (Groups A, C and E). This can be explained by the computation of hamming distance, which takes half more time to run.

The hash function therefore appears to be very fast when it is used with sequences having a size between 100 and 1000 bp (Groups A, B, C and D) and fewer slower with 10000pb sequences (Groups E, F). This is due to the complexity of the DCT function that is O(n * log (n)). In order to have more statistical representation, complementary sets of simulations would be perform with 100 million of simulations per group. At the moment, we did not perform these simulations because it would takes more than 80 days to run, on our testing computer.

### 4. RESULTS

The following graphs show the statistical distribution of Hamming Distances (X axis) for each class of divergence rate for 6 different groups of simulations. Each color represents a divergence rate. The Y axis represents the percentage rate of divergent classes. For each Hamming Distance number, it is possible to determine the proportion of divergence rate.

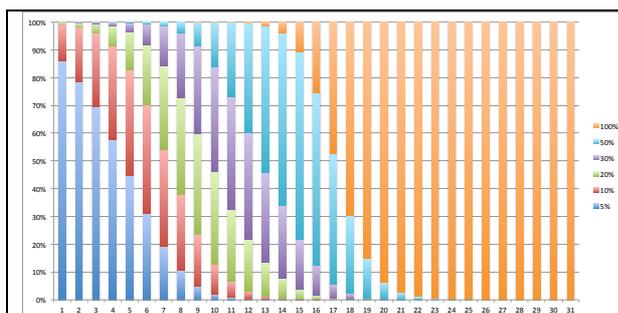

**Figure 9:** Statistical distribution of Hamming Distance with simulation of group A.

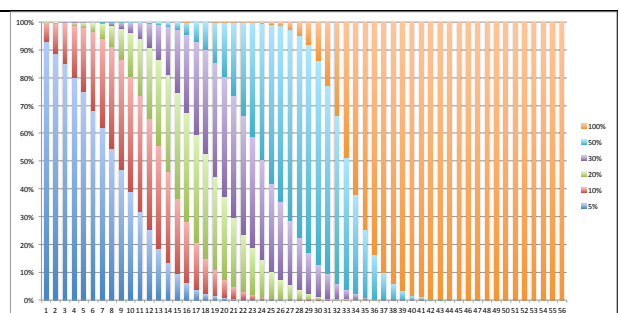

**Figure 10:** Statistical distribution of Hamming Distance with simulation of group B.

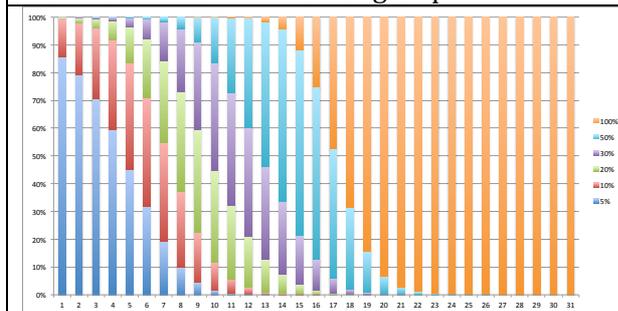

**Figure 11:** Statistical distribution of Hamming

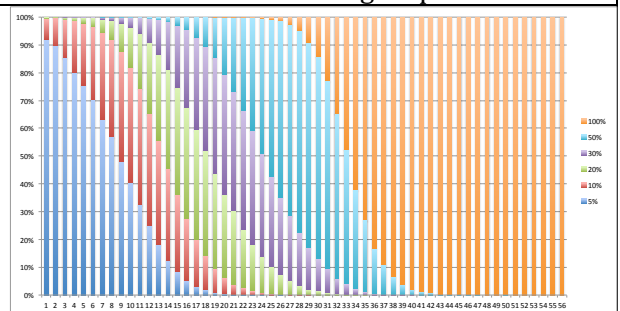

**Figure 12:** Statistical distribution of Hamming

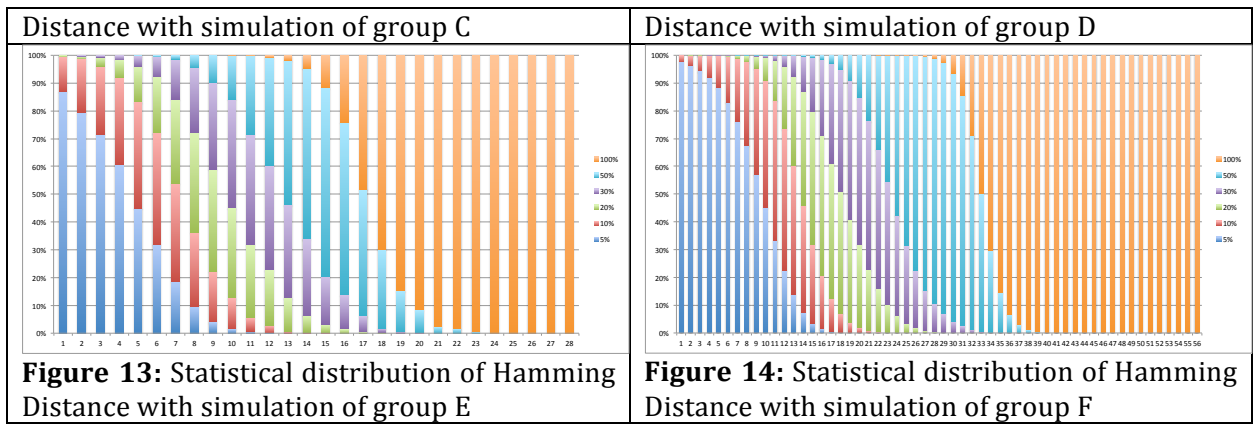

| Distance with simulation of group C | Distance with simulation of group D |
|---|---|
| **Figure 13:** Statistical distribution of Hamming Distance with simulation of group E | **Figure 14:** Statistical distribution of Hamming Distance with simulation of group F |

The distribution of Hamming Distances is close for each group of simulations that compare hashes of the same size. However, the length of sequences and the divergent rate do not seem to have a real impact. This can be explained by the fact that the coefficients used to generate the hashes are the most significant. However, a largest hash appears to provide a better sensitivity. Even if 64-bits hashes are more precise, the comparison of 32-bits hashes could be an acceptable similarity index, including for DNA sequences having a length of 10 000pb. In this case, the ratio between sequence and hash would be 10 000pb to 4 characters (2500x).

## 4. CONCLUSIONS AND FUTURE WORK

In this article, we have presented a new method to store and retrieve large scale, based on a perceptual hash function, using a DCT-SO. To our knowledge, this is the first approach of this type. The DNA sequence hashing methods described in the literature, generally refer to methods of string indexing or suffixes trees based algorithms.

In term of validation, a first evaluation by theoretical simulations has been performed. A large number of sequences having exactly the same length have been compared. Those theoretical simulations, demonstrated the speed and reliability of our hash function despite a drastically data reduction between originals sequences and hashes. The similarity index is calculated using the Hamming Distance. When two hashes are compared, this index does not strictly determine the degree of similarity but theoretical simulations show the efficiency of the approach for compressing sequence and measuring their similarities. So, we can conclude that it is a reliable indicator for inferring the proximity between two sequences.

We have seen through this article, the encoding DNA sequences in a matrix of pixels and the using of the hash function could be a quick and reliable method to index and compare DNA sequences. In the same way, it would be interesting to carry out new experiments on protein sequences. Thus, the 22 amino acids may be encoded as a gray scale matrix of pixels and could be hashed and compared with the same method.

Future tests would be validating the use of this hash function through real datasets. This is to evaluate if it could be a reliable method to identify reference genomes of sequences, which belongs to metagenomics sample. Reference genomes would be indexed, by using the hash function, to form a hash table stored into a database. The sequences to be compared usually have different length; therefore the indexing phase should be performed not on full sequences but on subsequences with a fixed size. The aim could be to return quickly, for a DNA sequence candidate, all reference sequences present in the database with high probability to be the nearest.

In terms of perspective, we plan to extend our work to sequence alignment and to compare it with distributed version of BLAST such as PLAST [9], but also to the work of Saldias and al. [10].